\documentclass[useAMS,usenatbib]{mn2e}

\usepackage{graphicx,epsfig}

\def\gsim{\;\lower4pt\hbox{${\buildrel\displaystyle >\over\sim}$}\;}
\def\lsim{\;\lower4pt\hbox{${\buildrel\displaystyle <\over\sim}$}\;}
\def\grls{\;\lower4pt\hbox{${\buildrel\displaystyle >\over <}$}\;}

\title[Cloud EECC Dynamics and Line Profiles]
{Global Collapses and Expansions in Star-Forming Clouds}
\author[Y. Gao \& Y.-Q. Lou]
{Yang Gao$^{1}$\thanks{
  E-mail: gaoyang-00@mails.tsinghua.edu.cn (YG);\hskip 1.5cm\hbox{}
  louyq@mail.tsinghua.edu.cn,\ \ lou@oddjob.uchicago.edu (Y-QL)}
  and Yu-Qing Lou$^{1,2,3}$\footnotemark[1]\\
$^1$Department of Physics and Tsinghua Centre for Astrophysics,
  Tsinghua University, Beijing 100084, China \\
$^2$National Astronomical Observatories of China, Chinese Academy
  of Sciences, A20, Datun Road, Beijing 100012, China \\
$^3$Department of Astronomy and Astrophysics,
  The University of Chicago, 5640 South Ellis Avenue, Chicago, IL 60637, USA }
\date{Accepted 2009 August 29. Received 2009 August 28;
      in original form 2008 December 9}
\begin{document}
\maketitle

\begin{abstract}
Spectral molecular line profile observations of star-forming
  molecular clouds sometimes show distinct red asymmetric
  double-peaked molecular line profiles with weaker blue
  peaks and stronger red peaks.
  For some star-forming molecular clouds, such molecular
  transitions with red asymmetric line profiles and blue
  asymmetric line profiles (i.e. blue asymmetric double-peaked
  molecular line profiles with weaker red peaks and stronger blue
  peaks) may coexist in spatially resolved spectral observations,
  while for others, such molecular transitions with red asymmetric
  line profiles may completely dominate in spatially resolved
  spectral observations. Blue asymmetric line profiles are usually
  interpreted as signals of central core collapses, while red
  asymmetric line profiles remain unexplained.
In this paper, we advance a spherically symmetric self-similar
  hydrodynamic model framework for envelope expansions with core
  collapses (EECC) of a general polytropic molecular gas cloud
  under self-gravity.
Based on such EECC hydrodynamic cloud models, we perform tracer
  molecular line profile calculations using the publicly
  available RATRAN code for star-forming clouds with
  spectroscopic signatures of red asymmetric line profiles.
The presence of red asymmetric line profiles from molecular cloud
  cores indicates that EECC processes are most likely an essential
  hydrodynamic process of star formation.
With spatial distributions, we explore various profiles of
  molecular lines for several tracer molecules in different
  settings of EECC dynamic models with and without shocks.
\end{abstract}



\section{Introduction}

Collapses, expansions, shocks and turbulence are several important
  dynamic features of star formation processes occurring inside
  molecular clouds. Such molecular cloud dynamic characteristics
  may be revealed by rich diagnostics and comprehensive analysis
  of molecular spectral emission line profiles.
An extensively discussed theoretical framework of forming low-mass
  stars is the `inside-out collapse' scenario \citep{shu1977,shu1987}.
  This model describes an isothermal self-similar dynamic solution
  that has a collapsing core surrounded by a static envelope with an
  expanding boundary engulfing more and more mass into the collapsed
  region.
By adopting empirically inferred temperature variations to
  replace the constant temperature, this dynamic collapse
  structure may lead to double-peak molecular line profiles with blue
  peaks stronger than red peaks (i.e. blue profiles), as revealed by
  spectral line observations of some molecular globules in early stages
  of star formation \citep[e.g.][]{zhou1993,saito1999,hogerheijde2000}.
Collapse solution based on dynamics of general polytropic gas
  sphere can also lead to blue profiles, yet with a temperature
  variation involved in the dynamic model in a self-consistent
  manner \citep[][]{gao2009}.


Clearly, a collapse model is not a full story of star formation,
  as there exist observational signatures that cannot be accounted for
  in the collapse scenario \citep[e.g.][]{wilner2000,belloche2002,vandertak2005}.
One important signature is the detected molecular emission lines
  with red asymmetry, i.e. optically thick emission lines are red
  shifted relative to optically thin lines from the same source.
A statistical survey of the observed optically thick molecular
  lines show that a quarter up to $\sim 30\%$ of all sources show
  red asymmetry \citep[e.g.][]{mardones1997,evans2003,fuller2005}.
  Among sources of red asymmetry, double-peak molecular line
  profiles with red peaks stronger than blue peaks (i.e. red profiles)
  are further identified \citep[e.g.][]{park2004,tafalla2006,velusamy2008}.
More specifically, for a number of known cloud sources, both red
  and blue asymmetries are observed towards the same transitions of
  the same molecules but at different beam offsets away from the
  centre \citep[e.g.][]{tafalla1998,matthews2006}.
While blue profiles are recognized as the signature of radial
  infalls or core collapses, what then do these red profiles imply?

Earlier radiative transfer calculations with parameterized flow
  structures indicate that rotation and radial outflows in molecular
  clouds may produce red profiles
  \citep[e.g.][]{adelson1988}.
Dynamics of bipolar outflows has been extensively studied in the
  past \citep[e.g.][]{shu1991,shu1994,fiege1996a,ostriker1997,matzner1999,shang2006},
  and their spectroscopic signatures (e.g. molecular line profiles and
  radio maps) are also explored \citep[e.g.][]{fiege1996b}.
Being widely observed, these molecular bipolar outflows have
  velocities $\gsim 10~{\rm km~s}^{-1}$ \citep[e.g.][]{wu2005,su2007},
  which are too large to account for the widely observed red asymmetry
  signatures indicating a typical gas flow velocity of
  $\sim 1~{\rm km~s}^{-1}$ \citep[e.g.][]{mardones1997,fuller2005}. Of
  course, for special cases when bipolar outflows are oriented close
  to the plane of the sky, molecular line splittings can be smaller.
Line profiles (as well as millimetre continuum maps) in a
  molecular cloud with bipolar outflow show an asymmetric spatial
  distribution
  according to the axis direction of the outflow
  \citep[e.g.][]{difrancesco2001,matthews2006,jorgensen2007}.
Red profiles caused by rotation around the central core have a
  systematic displacement in the velocity of local standard of rest
  ($V_{\rm {LSR}}$) \citep[][Redman et al. 2004]{difrancesco2001}.
The spatial distribution of red profiles is also asymmetric
  according to the direction of the rotation axis
  \citep{park1992,zhou1995}.
Though generally considered to be part of contributions to the
  line broadening, turbulence in clouds is sometimes invoked to form
  line profiles with red or blue asymmetries, especially those
  more violent ones
  in the outer layers of clouds
  \citep[e.g.][]{ossenkopf2002,lee2009}.
Contraction and expansion motions caused by large-scale thermal
  pulsations (with a typical oscillation period of
  $\sim 10^5$~yr) in starless cores can also be origins of
  asymmetric molecular line profiles from molecular clouds
  in their early stages of star formation, i.e. starless
  cores \citep[][]{lada2003,redman2006,keto2006,aguti2007}.

Theoretical models of envelope expansion with core collapse
 (EECC) for star formation represent a significant recent
 development \citep[][Yu \& Lou 2005; Yu et al. 2006;
 Hu \& Lou 2008]{lou2004,ShenLou2004,lou2006,wang2008}.
   Such EECC dynamic process
   might affect the initial mass function of stars
   (e.g. Nakano et al. 1995) and the environment of star formation
    \citep[e.g.][]{matzner2000,moraghan2008}.
    The crucial question is whether such global dynamic structures
  actually exist in star-forming clouds with sensible observational
  diagnostics such as characteristic features in molecular profiles?
  The main thrust of this paper is to show that theoretical molecular
  line profiles based on the general polytropic EECC solutions with
  collapses, expansions and shocks
  are able to explain observations of red profiles and can further
  provide plausible inferences of the star-forming region.

This paper is structured as follows.
  We first present possible self-similar dynamic structures
  of collapse and expansion solutions for general polytropic
  molecular clouds \citep{wang2008} in Section 2.
Molecular spectral line profiles are then obtained from
  radiative transfer calculations in Section 3.
Exploration of the EECC cloud conditions that generate
  red asymmetric line profiles and our perspective of
  model applications are also presented therein.
We summarize and conclude in Section 4.

\section{Collapse and Expansion Dynamics}

\subsection{General Polytropic Hydrodynamic Models}

To describe star-forming molecular clouds, we adopt the general
  polytropic self-similar model framework of \citet{wang2008}
  but without the random magnetic field. In spherical polar
  coordinates $(r,\ \theta,\ \phi)$, nonlinear hydrodynamic
  partial differential equations (PDEs) for spherically
  symmetric molecular cloud dynamics are
\begin{equation}
  {{\partial\rho}\over{\partial t}}
  +{1\over{r^2}}{{\partial}\over{\partial r}}(r^2\rho u)=0\ ,
  \label{Equ:mass1}
\end{equation}
\begin{equation}
  {{\partial u}\over{\partial t}}+u{{\partial u}\over {\partial
  r}}=-{1\over{\rho}}{{\partial p}\over {\partial
  r}}-{{GM}\over{r^2}}\ ,\label{Equ:force}
\end{equation}
\begin{equation}
  {{\partial M}\over{\partial t}}+u{{\partial M}\over{\partial
  r}}=0\ ,\qquad\qquad\qquad {{\partial M}\over{\partial r}}=4\pi
  r^{2}\rho\ ,\label{Equ:mass2}
\end{equation}
\begin{equation}
  \left(\frac{\partial}{\partial t}+u\frac{\partial}{\partial r}\right)
  {\rm ln}\left(\frac{p}{\rho^{\gamma}}\right)=0\ ,\label{Equ:entropy}
\end{equation}
  where mass density $\rho$, radial flow velocity $u$, thermal gas
  pressure $p$, and enclosed mass $M$ depend on radius $r$ and time $t$;
  $G=6.67\times 10^{-8}\hbox{ dyne cm}^2\hbox{ g}^{-2}$ is the
  gravitational constant and $\gamma$ is the polytropic index.
  Equations (\ref{Equ:mass1}) and (\ref{Equ:force}) are mass
  and radial momentum conservations, respectively; equation
  (\ref{Equ:mass2}) is another form of mass conservation.
Equation (\ref{Equ:entropy}) is the conservation of specific
  entropy along streamlines, which implies a general polytropic
  state equation (EoS) $p=K(r,\ t)\rho^{\gamma}$ with $K(r,\ t)$
  being a coefficient that varies with both time $t$ and radius $r$
  in general.

These nonlinear PDEs allow  self-similar solutions and the
  pertinent self-similar transformation is given below
\begin{eqnarray}
  r=k^{1/2}t^n x\ ,
  \label{equ:radius}
\end{eqnarray}
\begin{eqnarray}
  u=k^{1/2}t^{n-1}v(x)\ ,\qquad\qquad
  \rho=\frac{\alpha(x)}{4\pi Gt^2}\ ,
  \label{equ:varu}
\end{eqnarray}
\begin{eqnarray}
  M=\frac{k^{3/2}t^{3n-2}m(x)}{(3n-2)G}\ ,\qquad
  p=\frac{kt^{2n-4}\alpha(x)^{\gamma}m(x)^q}{4\pi G}\ ,\
  \label{equ:varp}
\end{eqnarray}
  where $x$ is the self-similar independent variable combining
  $r$ and $t$ in a special manner, and $\alpha(x)$, $m(x)$ and
  $v(x)$ are dimensionless reduced mass density, enclosed mass
  and radial flow velocity, respectively. According to
  mass conservation equation (\ref{Equ:mass2}), the reduced
  enclosed mass can be expressed as $m(x)=\alpha(x) x^2 [nx-v(x)]$.
Being an important thermodynamic variable, the gas temperature $T$
  is given by the ideal gas law
  \begin{equation}
  T\equiv\frac{p}{k_{\rm B}\rho/(\mu m_{\rm H})}=
  \frac{\mu m_{\rm H}}{k_B}kt^{2n-2}\alpha(x)^{\gamma-1}m(x)^q\ ,
  \label{equ:temp}
  \end{equation}
  where $k_{\rm B}$, $\mu$ and $m_{\rm H}$ are Boltzmann's constant,
  mean molecular weight and hydrogen mass, respectively.
We shall adopt $\mu\cong 1$ for typical star-forming clouds. For a
  finite $dm(x)/dx$ as $x\rightarrow 0$, equation ({\ref{equ:varp}})
  gives a central mass accretion rate
  \begin{equation}
  \dot{M_0}=k^{3/2}t^{3(n-1)}m_0/G\ ,\label{equ:massaccretion}
  \end{equation}
  with $m_0$ being the central reduced point mass enclosed.
  For $n=1$, the central mass accretion rate $\dot{M_0}$ remains
  constant; for $n>1$ and $n<1$, this $\dot{M_0}$ increases and
  decreases with increasing time, respectively.
Indices $\gamma$, $n$ and $q$ are related by
  general polytropic EoS (\ref{Equ:entropy}) with
  $q=2(n+\gamma-2)/(3n-2)$. The case $n+\gamma=2$ features an
  EoS for a conventional polytropic gas and $\gamma=n=1$
  describes an isothermal gas. Self-similar transformation
  equations (\ref{equ:radius})$-$(\ref{equ:varp}) make it
  possible to cast nonlinear PDEs (\ref{Equ:mass1})$-$(\ref{Equ:entropy})
  into nonlinear ordinary differential equations (ODEs) in terms
  of $x$ which can be solved numerically with analytical asymptotic
  conditions and by taking care of the sonic critical curve (see Wang
  \& Lou 2008 for details).

Analytically, we have a static equilibrium solution for spherical
cloud, namely, a singular polytropic sphere (SPS)
\begin{eqnarray}
v=0\ ,\quad\
\alpha=\bigg[\frac{n^{2-q}}{2(2-n)(3n-2)}\bigg]^{-1/(n-3nq/2)}
x^{-{2}/{n}}\ ,\nonumber\\
\!\!\!\!\!\!\!\!\!\!\!\!\!\!\!\!\!\!\!\!\!\!\!\!\!\!\!
m=n\bigg[\frac{n^{2-q}}{2(2-n)(3n-2)}\bigg]^{-1/(n-3nq/2)}
x^{(3n-2)/n}\quad\label{Equ:static}
\end{eqnarray}
(e.g. Lou \& Hu 2009). This static SPS solution may be helpful for
speculating the origin of those self-similar collapse and
expansion solutions (see below).


In the limit of $x\rightarrow +\infty$, we have asymptotic
similarity solution to the leading orders, viz.
\begin{eqnarray}
\alpha=Ax^{-{2}/{n}},
\qquad\qquad\qquad\qquad\qquad\qquad\qquad\qquad
\qquad \nonumber \\
v=\bigg[-\frac{nA}{(3n-2)}+2(2-n)n^{q-1}A^{1-n+3nq/2}\bigg]
x^{{(n-2)}/{n}}\nonumber\\
+Bx^{{(n-1)}/{n}}\ ,\label{Equ:infinity}
\end{eqnarray}
where $A$ and $B$ are two constants of integration, referred to as
the mass and velocity parameters, respectively.
In the other limit of $x\rightarrow 0^{+}$, the asymptotic central
free-fall solution is
\begin{equation}
v=-\bigg[\frac{2m_{0}}{(3n-2)x}\bigg]^{1/2}\ , \qquad\quad
\alpha=\bigg[\frac{(3n-2)m_{0}}{2x^{3}}\bigg]^{1/2}\ ,
\label{Equ:zero1}
\end{equation}
where $m=m_0$ is the reduced enclosed mass $m(x)=\alpha x^2(nx-v)$
as $x\rightarrow 0^{+}$ for a point mass at the very centre.

\subsection{General Polytropic EECC Solutions}


We mainly focus on self-similar dynamic solutions for
  envelope expansions with core collapse in the form of free-fall
  towards the cloud centre (i.e. EECC solutions; Lou \& Shen 2004
  and Shen \& Lou 2004).
Relevant parameters of five such selected self-similar solutions
  are listed in Table \ref{Table:dynamics}, where $x_{\rm inf}$
  is the outgoing boundary separating the collapse and expansion
  regions; $m_0$ is the dimensionless central reduced point mass;
  and $m_{\rm tot}$ is the total reduced enclosed mass of a spherical
  cloud with an expanding outer edge at $x=5$ for the star-forming
  cloud. We also list in Table \ref{Table:dynamics} two important
  mass and velocity parameters $A$ and $B$, which characterize
  asymptotic dynamic behaviours for $x\rightarrow +\infty$ and serve
  as asymptotic `boundary' conditions in constructing general
  polytropic EECC solutions (see eqs (26) and (27) of Wang \& Lou
  2008 for details).
Solutions IV and V are two EECC shock solutions with $\gamma=1.2$
  and $n=0.8$, the upstream point $x_1$ and downstream point $x_2$
  correspond to the same shock radius $R_{\rm sh}$; $v_1$ and $v_2$
  are the upstream and downstream reduced radial flow velocities,
  with negative value being infall and positive value
  being expansion, respectively.

\begin{table*}
 \centering
 \begin{minipage}{160mm}
 \begin{center}
  \begin{tabular}{@{}lllllllllllll@{}}
  \hline
 ${\rm No.}$&$\gamma$&$n$&$q$&$x_{\rm inf}$&$m_0$&$m_{\rm tot}$&$A$&$B$&$x_1$&$v_1$&$x_2$&$v_2$\\
 \hline
 I & 1.1 & 0.8 & $-1/2$ & 1.50 & 1.49 & 5.0 & 3.5 & 2.2 &---&---&---&--- \\
 II & 1.2 & 0.8 & 0 & 1.77 & 2.32 & 10.0 & 5.0 & 2.4 &---&---&---&---\\
 III & 1.2 & 0.9 & 2/7 & 1.80 & 3.74 & 20.8 & 10.0 & 2.4 &---&---&---&---\\
 IV & 1.2 & 0.8 & 0  & 0.51 & 0.207 & 4.6 & 2.68 & 0.47 & 2.77 & $0.01$ & 2.57 & 1.54 \\
 V & 1.2 & 0.8 & 0  & 0.10 & 0.0254 & 9.2 & 7.72 & 3.70 & 1.70 & $ -0.30 $ & 1.68 & 0.58 \\
\hline
\end{tabular}
\end{center}
  \caption{Parameters of self-similar EECC dynamic solutions
  without and with shocks for five Models I$-$V labelled by
  roman numerals on the left most column.
  Among the three scaling indices $\gamma$, $n$ and $q$, only
  two are independent and all three are related by general
  polytropic EoS (\ref{Equ:entropy}) with
  $q=2(n+\gamma-2)/(3n-2)$. Three parameters
  $x_{\rm inf}$, $m_0$ and $m_{\rm tot}$ are the dimensionless
  infall radius separating the inner collapse and outer
  expansion regions, the reduced central point mass, and the
  reduced total enclosed mass for a model cloud, respectively.
Two coefficients $A$ and $B$ are mass and velocity parameters in
  asymptotic  `boundary' conditions in constructing corresponding
  self-similar solutions as $x\rightarrow +\infty$.
Parameters $x_1$, $x_2$, $v_1$ and $v_2$ are the upstream and down
  stream locations and velocities for solutions across an outgoing
  shock front (i.e. Models IV and V).
  \label{Table:dynamics}}
\end{minipage}
\end{table*}

Each of these selected EECC solutions has
  a central collapsed core and an outer expansion envelope with
  an outgoing interface at $x_{\rm inf}$ separating the two zones.
At the beginning of evolution ($t\rightarrow 0^+$), this boundary
  radius is approximately zero according to self-similar
  transformation (\ref{equ:radius}), which indicates that
  collapse begins from the very centre of the cloud.
As time goes on, this boundary radius increases with a constant
  $x_{\rm inf}$ at a variable speed of $u_b=nk^{1/2}t^{n-1}x_{\rm{inf}}$
  according to equation (\ref{equ:radius}), which means that
  more and more mass are enclosed into the collapse region
  (see expression (\ref{equ:varp})).
At a given time $t$, mass densities and gas temperatures of all
  these EECC solutions increase towards the cloud centre.
Shock solutions involve discontinuities of flow velocities,
  densities and temperatures across shock radius $R_{\rm sh}$.
  Their dynamic structures are shown in Figs. \ref{fig:physical1}
  and \ref{fig:physical2} by adopting estimated physical scalings
  described below.

In constructing self-similar hydrodynamic solutions, we have
  presumed a priori the possible existence of such form of similarity
  solutions under plausible asymptotic conditions and actually derive
  them analytically and/or numerically from nonlinear hydrodynamic
  equations satisfying relevant physical constraints. Meanwhile, it
  is of considerable interest to figure out even qualitatively how a
  molecular cloud system can sensibly evolve into such a self-similar
  phase given a certain class of initial and boundary conditions. This
  is a challenge especially in view of the existence of several possible
  asymptotic self-similar solutions and requires a deeper theoretical
  understanding (e.g. Wang \& Lou 2008). At this stage, we tentatively
  offer speculations on possible scenarios leading to self-similar EECC
  dynamic evolution invoked in this paper for modelling the dynamics of
  certain star-forming molecular clouds.

Similar to stellar oscillations widely studied observationally
   and theoretically, molecular clouds in SPS equilibrium
   (\ref{Equ:static}) when somehow perturbed may give rise to
   acoustic pulsations on much larger spatial and temporal scales
   \citep[e.g.][]{lada2003,redman2006,keto2006,aguti2007}.
For simplicity, we may envision purely radial acoustic pulsations
   with possible radial nodes in spherical molecular clouds; for
   example, such radial pulsations might be induced or excited by
   a sufficiently massive companion or transient object.
With idealizations, such acoustic pulsations might persist
   periodically for a long time in molecular clouds.
Realistically, such acoustic pulsations might be `damped' in one
   or two `periods' due to radiative losses as well as nonlinear
   effects. Among various pulsation phases, it would be possible
   to have a phase characterized by core contractions with
   envelope expansions.
With such `initial' conditions in molecular clouds, the nonlinear
  evolution may eventually lead to core collapse under the
  self-gravity while the envelope expands into the surrounding
  interstellar medium (ISM). It is emphasized that no pulsations
  are necessarily persistent in this scenario. We speculate that
  this might evolve into self-similar EECC dynamic phase advanced
  in this paper. Following this scenario, pulsations of molecular
  clouds with different phases may evolve nonlinearly into a
  variety of dynamic states.

The EECC solution is necessarily consistent with the aspect of
  energy conservation. For most low-mass star formations, the
  Kelvin-Helmholz time scale $t_{\rm KH}=GM_{\rm tot}^2/(RL)$
  is longer than the dynamic timescale $(\rho G)^{-1/2}$,
  because of lower radiative efficiency \citep[see e.g.][]{mckee2007}.
Here $M_{\rm tot}$, $R$ and $L$ are the total mass, the outer radius
  and the luminosity of the cloud, respectively.
Because of radiative inefficiency, most low-mass star forming
  clouds need an extra means to carry out gravitational energy
  that are released during the central accretion. Global envelope
  expansion, which is a sensible dynamic solution as already shown,
  serves as the extra energy release.
From the consideration of energy conservation, the EECC shock
  solutions may potentially offer valuable clues to the problem
  that accretion rates derived from observed luminosity are much
  smaller than those expected according to their dynamic evolution
  (i.e. the luminosity problem) in low-mass star formations
  \citep[e.g.][]{kenyon1990,mckee2007}.
  For the physical scenario of variable central mass accretion
  rate and thus variable luminosity for the formation of
  low-mass stars in molecular clouds, we provide a general
  polytropic model explain the `luminosity problem' (Lou \& Dong
  2009 in preparation).
%

\subsection{Physical Properties of Molecular Clouds}

The reduced dynamic variables should be converted to
  physical variables as applied to realistic cloud systems.

The typical infall radius of a molecular cloud is $\sim 0.01 -
  0.03$~pc \citep[e.g.][]{myers2005}, or $\sim 10^3 - 10^4$~AU.
As the reduced infall radius is at $x\sim 1$, we may choose
  the length scale as
  \begin{equation}
  k^{1/2}t^n\sim 4\times10^3~{\rm AU}
  \label{equ:lengthscale}
  \end{equation}
  in self-similar transformation (\ref{equ:radius})$-$(\ref{equ:varp}).
  The outer cloud radius is set at $R\sim 2\times 10^4$ AU.
The number density at the infall radius is estimated by
  $\sim 10^4-10^5~{\rm cm}^{-3}$ \citep[e.g.][]{harvey2003,evans2009}.
A reduced number density of unity implies
  \begin{equation}
  (4\pi G\mu m_{\rm H}t^2)^{-1}\cong 9\times10^4~{\rm cm}^{-3}\ ,
  \label{equ:densityscale}
  \end{equation}
leading to an estimated dynamic timescale. With parameter scalings
  (\ref{equ:lengthscale}) and (\ref{equ:densityscale}) for clouds,
  physical variables of a cloud can be expressed as follows according
  to equations (\ref{equ:radius})$-$(\ref{equ:massaccretion}), namely
  \begin{equation}
  r=4\times 10^3\ x~{\rm AU}\ ,
  \label{equ:realradius}
  \end{equation}
  \begin{equation}
  u=0.213\ v(x)~{\rm km~s^{-1}}\ ,
  \label{equ:realvelo}
  \end{equation}
  \begin{equation}
  N=9\times10^4\ \alpha(x)~{\rm cm^{-3}}\ ,
  \label{equ:realNdensity}
  \end{equation}
  \begin{equation}
  M=0.204\ m(x)/(3n-2)~M_\odot\ ,
  \label{equ:realmass}
  \end{equation}
  \begin{equation}
  T=5.33\ \alpha(x)^{\gamma-1}m(x)^q~{\rm K}\ ,
  \label{equ:realtemp}
  \end{equation}
  \begin{equation}
  \dot{M}_0=2.26\times10^{-6} m_0~M_\odot~{\rm yr}^{-1}\ ,
  \label{equ:realmassacc}
  \end{equation}
  where $N=\rho/(\mu m_{\rm H})$ is the particle number density.
We note that the dynamic timescale of a cloud is also
  automatically fixed, i.e. $t_d\sim 2.2\times 10^5~$yr
  according to scaling estimate (\ref{equ:densityscale}).
From scaling estimates (\ref{equ:lengthscale}) and
  (\ref{equ:densityscale}), the sound parameter $k$ is
  estimated by $k^{1/2}=3.76~{\rm km~s^{-0.9}}$ for
  solution III and by $k^{1/2}=65.7~{\rm km~s^{-0.8}}$ for
  solutions I, II, IV and V;
  while the upstream $k^{1/2}$
  jumps up to $k^{1/2}=70.8~{\rm km~s^{-0.8}}$ and
  $k^{1/2}=66.5~{\rm km~s^{-0.8}}$ on the downstream
  side of shock solutions IV and V, respectively.

\begin{table*}
 \centering
 \begin{minipage}{180mm}
 \begin{center}
 \begin{tabular}{@{}lllllllllllll@{}}
  \hline
 {\rm No.} & $\gamma$ & $n$ & $M_0$ & $M_{\rm tot}$ & $\dot{M}_0$ ($M_\odot~{\rm yr}^{-1}$) & $ R_{\rm inf}\ ({\rm AU})$ & $R_{\rm sh}\ ({\rm AU})$ & $u_1$ & $u_2$\\
 \hline
 I & 1.1 & 0.8 & $0.75~M_\odot$ & $2.52~M_\odot$ & $3.38\times10^{-6}$ & $ 6.0\times10^3 $  & --- & --- & ---  \\
 II & 1.2 & 0.8 & $1.18~M_\odot$ & $5.10~M_\odot$ & $5.30\times10^{-6}$ & $ 7.1\times10^3 $ & --- & --- & --- \\
 III & 1.2 & 0.9 & $1.91~M_\odot$ & $11.00~M_\odot$ & $8.41 \times10^{-6}$ & $ 7.2\times10^3$ & --- & --- & --- \\
 IV & 1.2 & 0.8 & $0.106~M_\odot$ & $2.34~M_\odot$ & $4.73\times10^{-7}$ & $ 2.04\times10^3 $ & $11.1\times10^3$& $0.002~{\rm km~s^{-1}}$ & $0.34~{\rm km~s^{-1}}$ \\
 V & 1.2 & 0.8 & $0.012~M_\odot$ & $4.68~M_\odot$ & $5.77\times10^{-8}$ & $ 0.40\times10^3  $ & $6.8\times10^3$ & $-0.07~{\rm km~s^{-1}}$ & $0.13~{\rm km~s^{-1}}$ \\
 \hline
\end{tabular}
\end{center}
  \caption{Physical parameters of cloud Models I$-$V,
    with $\gamma$ and $n$ being their polytropic indices
    and scaling indices, respectively. Two dimensional masses
  $M_0$ and $M_{\rm tot}$ are the central point mass
    and the total cloud mass inside $R=2\times10^4~{\rm AU}$,
    while $\dot{M}_0$ denotes the central mass accretion rate.
  Being the infall radius, $R_{\rm inf}$ is the boundary between
    the infall and outflow regions. Parameter
  $R_{\rm sh}$ is the outgoing shock radius for shock solutions
  in Models IV and V, and $u_1$ and $u_2$ are the upstream and
  downstream radial flow velocities, with negative values
  denoting inflows.
  \label{Table:physical}}
\end{minipage}
\end{table*}

\begin{figure}
\begin{center}
\epsfig{figure=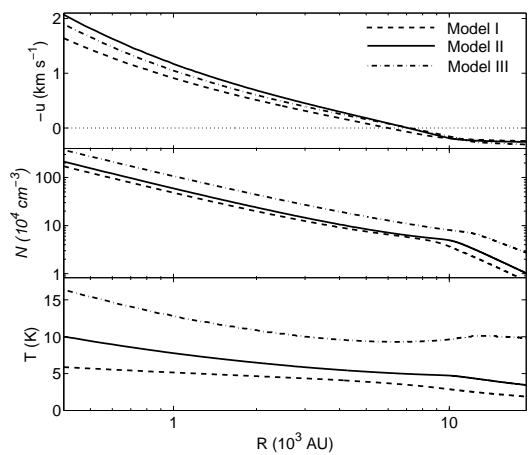,width=9.0cm,clip=}
\end{center}
\caption{Radial profiles for physical variables of three EECC
  dynamic models I, II and III, with different polytropic
  indices $\gamma$ and $n$ yet without shocks. From
top to bottom are: radial flow velocity (positive values for
  infall), number density and temperature profiles.
  The abscissa is radius $R$ in $10^3$ AU in a logarithmic scale.
Dashed, solid and dash-dotted curves are the EECC dynamic
  solutions for Model I with $\gamma=1.1$ and $n=0.8$,
  Model II with $\gamma=1.2$ and $n=0.8$, and Model III with
  $\gamma=1.2$ and $n=0.9$, respectively. The dotted
horizontal line in the top panel is for the zero
  velocity line $u=0~{\rm km~s^{-1}}$. The infall radii for
Models I, II and III are $R_{\rm
  inf}=6.0\times 10^3$ AU, $R_{\rm inf}=7.1\times 10^3$ AU
  and $R_{\rm inf}=7.2\times 10^3$ AU, respectively.
Other parameters for these general polytropic self-similar
  solutions are summarized in Tables \ref{Table:dynamics}
  and \ref{Table:physical}. \label{fig:physical1}}
\end{figure}

\begin{figure}
\begin{center}
\epsfig{figure=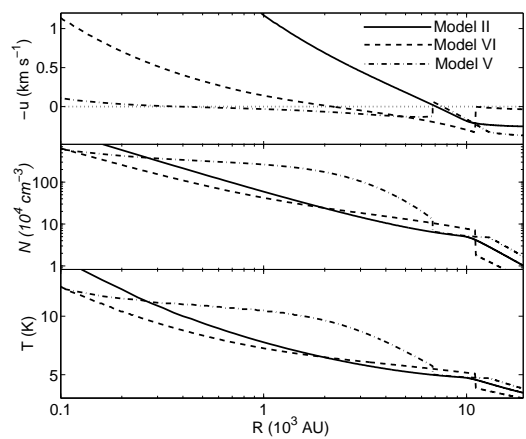,width=9.0cm,clip=}
\end{center}
\caption{Radial profiles for physical variables of three EECC
  dynamic Models II, IV and V, with and without shocks for
  polytropic index $\gamma=1.2$ and scaling index $n=0.8$.
  From top to bottom are: radial flow velocity (positive values for
  infall), number density and temperature profiles. The abscissa is
  radius $R$ in $10^3$ AU in a logarithmic scale.
Solid, dashed and dash-dotted curves are the solutions for Model II
  without a shock, Model IV with a shock at
  $R_{\rm sh}=11.1\times 10^3~{\rm AU}$,
  and Model V with a shock at
  $R_{\rm sh}=6.8\times 10^3~{\rm AU}$,
  respectively. The horizontal dotted line in the top panel
  is for the zero velocity line $u=0~{\rm km~s^{-1}}$.
The infall radii for Models II, IV and V are $R_{\rm
  inf}=7.1\times 10^3$ AU, $R_{\rm inf}=2.04\times 10^3$ AU
  and $R_{\rm inf}=0.40\times 10^3$ AU, respectively.
Other parameters for these self-similar EECC solutions are
  summarized in Tables \ref{Table:dynamics} and
  \ref{Table:physical}.\label{fig:physical2}}
\end{figure}

With scaling expressions (\ref{equ:lengthscale}) and
  (\ref{equ:densityscale}), we estimate physical properties
  of star-forming clouds as exemplified by Models I through
  V tabulated in Table \ref{Table:physical}.
Figures \ref{fig:physical1} and \ref{fig:physical2} illustrate
  the radial profiles for physical variables of these model
  clouds in two sets.

For those Models without shocks (i.e. I, II and III), the infall
  radius $R_{\rm inf}$ are similar to each other, and their central
  core mass $M_0$ and total mass $M_{\rm tot}$ are also comparable.
Here, the infall radius $R_{\rm inf}$ may be fairly small at the
  onset of a cloud core collapse and expands to encompass more gas
  and dust particles into the collapsed region as time goes on.
We note in Fig. \ref{fig:physical1} that the overall number
  density and temperature values increase from Model I to III;
  and the temperature of Model III increases slightly outwards at
  large radii ($\sim 10^4~{\rm AU}$).
We note in Table \ref{Table:physical} a gradual increase
  of the two masses $M_0$ and $M_{\rm tot}$ from Model I to
  Model III, caused by the increasing mass parameter $A$ as
  shown in Table \ref{Table:dynamics}
  [see equation (26) in \citep{wang2008}].
Our numerical explorations reveal that as $n+\gamma$ increases,
  the presence of EECC solutions calls for increasing values of $A$.
Another variation is that the ratio of central mass point to the
  total mass $M_0/M_{\rm tot}$ decreases from $\sim $29.9\% of Model I to
  $\sim $23.2\% of Model II to $\sim $17.4\% of Model III. This phenomena
  may be explained by accounting for general polytropic SPS solution
  (\ref{Equ:static}), of which smaller $\gamma$ and $n$ values
  lead to mass distributions with more central concentration.

By comparing Models II, IV and V, we can infer the effects of
  shocks in self-similar EECC dynamic solutions. From Fig.
  \ref{fig:physical2}, we see that central infall velocities for
  EECC solutions with shocks (i.e. Models IV and V) are greatly suppressed,
  and their infall radii $R_{\rm inf}$ are smaller as compared to
  Model II without shock (Table \ref{Table:physical}). As a consequence,
  the central
  mass accretion rate $\dot{M}_0$ and the central mass $M_0$ for Models
  IV and V are smaller, taking only 10\% and 1\% the values of Model II,
  respectively. The low values of mass accretion rates and central masses
  make these shock solutions potentially applicable for the formation of
  brown dwarfs. We should also note from Table \ref{Table:physical} that
  the total mass $M_{\rm tot}$ for these three models are comparable,
  which suggests the possibility that under certain conditions protostars
  with different masses may form from molecular clouds with similar masses.
In other words, molecular clouds of comparable masses have
  additional freedoms to give rise to central protostars of
  different masses.
The very low masses and mass accretion rates of EECC shock
  solutions are caused by efficient envelope expansions with small
  inner boundary radius (Table \ref{Table:physical}). Inside the shock
  radius $R_{\rm sh}$, the particle number density $N$ and gas temperature
  $T$ of Models IV and V are higher than or close to those of Model II
  (see Figure \ref{fig:physical2}).


From Table \ref{Table:physical}, we get the information that the
  central mass of a cloud (i.e. protostellar mass) is not tightly
  or directly related to the total cloud mass for different model
  solutions, especially for EECC dynamic solutions with shocks.
This conclusion agrees with the result that the star forming
  efficiency $M_0/M_{\rm tot}$ is mainly determined by the
  comparison between mass accretion and outflow rates (e.g. bipolar
  outflows therein), and is weakly dependent on the core
  mass\footnote{We use the two terms `core mass' and `cloud mass'
  interchangeably in the loose sense.} $M_{\rm tot}$ and number
  density $N$ \citep[e.g.][]{nakano1995,matzner2000}.
However, a constant ratio between mass outflow and accretion rates
  is always presumed for a certain multi-star-forming cloud, and the
  conclusion that initial mass function (IMF) of protostars should
  be closely connected with the cloud mass function (CMF) is usually
  drawn \citep[e.g.][]{hennebelle2008,myers2008}.
But there is no obvious reason that protostars forming in the same
  cloud should have the same mass outflow ratio over accretion.
Based on our analysis and results, we strongly suggest that
  solutions with different ratios between outflow and accretion
  exist in the same molecular cloud, and IMF of protostars can
  differ from CMF significantly, especially for those protostars
  with extremely low masses (i.e. brown dwarfs).

\begin{figure*}
\begin{center}
\includegraphics[width=\textwidth]{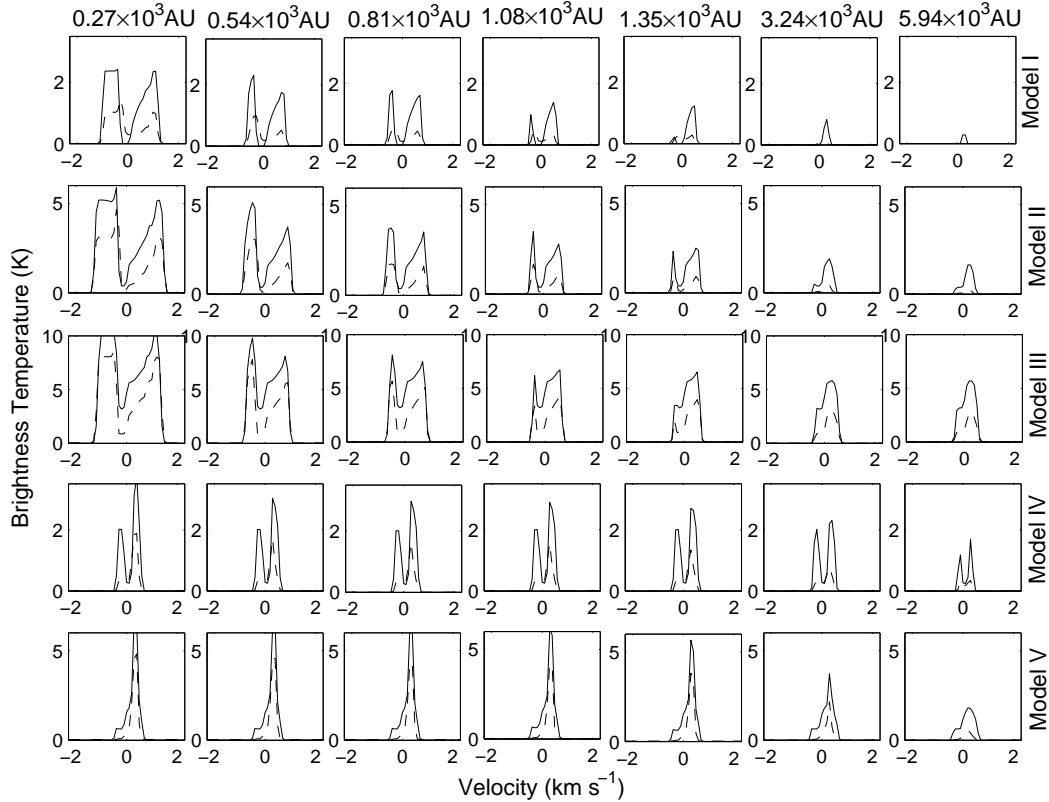}
\end{center}
\caption{Computed molecular line profiles of HCO$^+(1-0)$ at
  89.19~GHz (solid curves) and HCO$^+(3-2)$ at 267.56~GHz
  (dashed curves) from individual pixels of spatially resolved
  emission lines for five EECC dynamic models, viz. Models I--V.
The ordinate is the brightness temperature in Kelvin and the
  abscissa is the projected velocity component along the LOS in
  the local standard of rest (LSR) in unit of ${\rm km~s^{-1}}$.
Rows from top to bottom show molecular line profiles for different
  dynamic Models I$-$V, respectively.
Panels from left to right present LOS with seven different
  impact parameter $b$ (i.e. $0.27\times10^3~{\rm AU}$,
  $0.54\times10^3~{\rm AU}$, $0.81\times10^3~{\rm AU}$,
  $1.08\times10^3~{\rm AU}$, $1.35\times10^3~{\rm AU}$,
  $3.24\times10^3~{\rm AU}$, $5.94\times10^3~{\rm AU}$).
Impact parameters for pointings are not uniformly chosen on
purpose such that the transitions between blue profiles and red
profiles can be clearly identified for Models I, II and III;
meanwhile, the spatial coverage is sufficiently large to show
variations of molecular line profiles.
Containing both turbulence and thermal effects, the intrinsic line
broadening is chosen as $\Delta u=0.17~{\rm km}~{\rm s}^{-1}$.
\label{fig:spectra1}}
\end{figure*}



\section{Spectral Line Profile Signatures}

We perform radiative transfer calculations for molecular spectral
  line profiles using the publicly available numerical code RATRAN
  \citep{hogerheijde2000}
  under spherical symmetry. This RATRAN code has been benchmarked
  among seven other radiative transfer codes by Van Zadelhoff et al.
  (2002) against examples of star forming clouds like B335. Based on
  the Monte Carlo method \citep[e.g.][]{HH1964,Shreider1966,Bernes1979},
  this RATRAN code deals with both radiative transfer and non-local
  thermal equilibrium (non-LTE) excitations of atomic and molecular
  lines, making it readily adaptable to realistic astrophysical cloud
  systems.\footnote{For the purpose of testing and checking, our simple
  model radiative transfer calculations (Gao, Lou \& Wu 2009) are also
  performed by adjusting optical depths in parallel to the RATRAN code
  calculations for samples of red profiles. The relevant results are
  comparable.}
  We adopt dynamic cloud models for line profile calculations.
The dynamic and thermal parameters, namely radial flow velocity
  $u$, number density $N$ and gas kinetic temperature $T$ are obtained
  consistently from the general polytropic EECC dynamic Models I--V.
  The temperature of dusts in a cloud is assumed to follow the gas
  kinetic temperature $T$.
From the cloud centre to its outer radius $R=20000~{\rm AU}$, a
  molecular cloud is divided into 12 shells with enough accuracy
  for calculations (8 and 10 shells are also tested separately but
  with little variance in line profiles).
However, physical properties of these shells are not uniform in
  order to show the transition between infall and outflow for all
  the dynamic models.

In Figs. \ref{fig:spectra1}, \ref{fig:spectra2} and
  \ref{fig:spectra3}, we show a sample of computed molecular line
  profiles, viz. HCO$^+$ J$=1-0$ at 89.19~GHz and HCO$^+$ J$=3-2$
  at 267.56~GHz, for the five dynamic EECC solution Models I$-$V.
Molecular line profiles for transitions CO J$=2-1$ at 230.54~GHz,
  C$^{18}$O J$=1-0$ at 109.78~GHz, CS J$=2-1$ at 97.98~GHz and N$_2$H$^+$
  J$=1-0$ at 93.13~GHz are shown in Figs. \ref{fig:spectra4} and
  \ref{fig:spectra5}. All molecular data are obtained from the
  Leiden Atomic and Molecular Database \citep{schoier2005}.
Emission line profiles with different impact parameters $b$
  [i.e. distance of line of sight (LOS) from the cloud centre]
  and under different conditions of micro
  turbulence are shown in these figures.
More compressed separations between impact parameters for
  pointings are chosen for LOS passing through the most inner
  regions of clouds, in order to clearly show the transitions
  between blue profiles and red profiles for Models I$-$III.
The number density of HCO$^+$ molecules is assumed to be
  proportional to the overall molecular number density with
  a constant ratio, i.e. $N_{\rm HCO^+}=2\times 10^{-9}N$.
Molecules in Figs. \ref{fig:spectra4} and \ref{fig:spectra5} are
also assumed to be of constant abundance ratios, viz. $N_{\rm
CO}=5\times10^{-5}N$, $N_{\rm C^{18}O}=1\times10^{-7}N$, $N_{\rm
CS}=3\times10^{-9}N$ and $N_{\rm N_2H^+}=1.5\times10^{-10}N$,
respectively. All these molecular abundances are chosen according
to \citet{tafalla2006}, and the central abundance hole derived
from fitting procedure therein is not assumed in our model
calculations. Strictly speaking, for all molecules, the variation
of abundance ratio should be seriously taken into account
\citep{rawlings2001,tsamis2008}, and the constant ratio abundance
adopted here is just a first order approximation.

\subsection{Red Profiles from EECC Dynamic Models}

\begin{figure*}
\begin{center}
\epsfig{figure=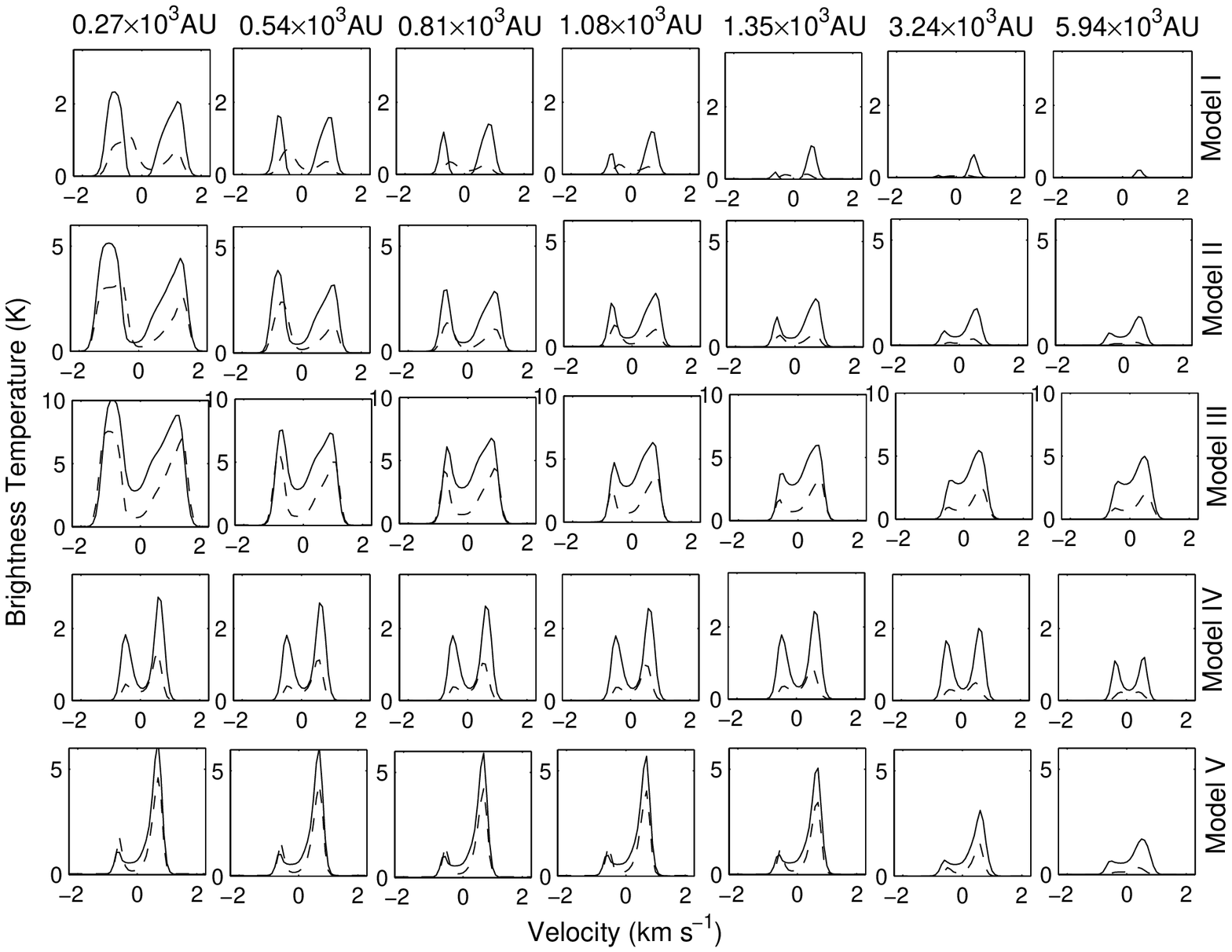,width=18.0cm,clip=}
\end{center}
\caption{Similar RATRAN computations for molecular spectra
  line profiles for five EECC dynamic Models I$-$V as in Fig.
  \ref{fig:spectra1}, but with intrinsic line broadening
  $\Delta u=0.5~{\rm km}~{\rm s}^{-1}$. This intrinsic line
  broadening contains both turbulence and thermal effects.
 \label{fig:spectra2}}
\end{figure*}

The most distinct spectroscopic signature of EECC dynamic models
  with core collapses and envelope expansions is the double-peak
  molecular line profile with the red peak being stronger than
  the blue peak,
  referred to as the red profile.
We readily see such red profiles from cloud models described by
  each EECC solution in Figs. \ref{fig:spectra1}--\ref{fig:spectra3}.
A more generic type of red profiles has no obvious central dip but
  shows a stronger red shoulder, which exist over a larger
  range of impact parameter $b$ for the LOS.
In Figs. \ref{fig:spectra4} and \ref{fig:spectra5}, red profiles
   also widely exist except for CO and C$^{18}$O transitions in
   Model III (Fig. \ref{fig:spectra4}).

We demonstrate two types of self-similar EECC dynamic solutions
  according to the appearance of red profiles for molecular spectral
  lines from star-forming clouds. For Models I, II and III, red
  profiles only exist for a fairly large impact parameter $b$, i.e.
  in Fig. \ref{fig:spectra1},
  $b\gsim 1.08\times10^3~{\rm AU}$ for Models I and III, and
  $b\gsim 1.35\times10^3~{\rm AU}$ for Model II;
  while inside these radii (i.e. impact parameter), blue profiles
  for molecular lines emerge because of the more dominant role of
  core collapse towards the centre.
This type of spatial transitions from blue profiles to red
  profiles can be seen in some molecular clouds with spatially
  resolved observations \citep[e.g.][]{tafalla2000,ward2001},
  although not sufficiently spherically symmetric.
For Models IV and V, whose infall radii are very small at
  the chosen epochs, red profiles for molecular lines exist
  throughout pixels with a range of impact parameter $b$;
  this can also be seen in some of recent observations
  \citep[][]{thompson2004,aguti2007}.
These results of our numerical exploration clearly indicate
  that the broad existence of red profiles is a characteristic
  signature of global envelope expansion, in contrast to the
  blue profiles which are characteristic feature of core
  collapse \citep[e.g.][]{zhou1993,gao2009}.
For all these EECC dynamic models without or with shocks,
  molecular emission line profiles decrease in the
  overall magnitudes as the LOS departs away from the
  core centre and gradually disappear as cloud density
  and temperature decrease further.

By comparing molecular line profiles among Models I, II and III,
  we find slight differences caused by different EECC dynamic
  profiles.
The higher spectral amplitude in Model III appears to be caused
  by the higher number density and gas kinetic temperature
  involved (Fig. \ref{fig:physical1}).
The central dips for line profiles of Model III are shallower
  than those from Models I and II; this is probably because
  of the gradual temperature rise at outer radii of Model III
  (Fig. \ref{fig:physical1}).
However, we cannot immediately claim that these differences in
  molecular line profiles are caused by different values of
  polytropic index $\gamma$ and scaling index $n$ in dynamic
  models, as for the same set of indices, different asymptotic
  boundary conditions will also lead to considerable variations
  in dynamic profiles \citep[see e.g.][]{lou2006,wang2008,HuLou2008}.
For Models IV and V, molecular emission lines with red profiles
  present for all values of impact parameter $b$. Then how could
  we understand the role of expanding shocks in these self-similar
  EECC dynamic solutions?
Dynamically, in gas clouds with relatively smooth or less drastic
  mass density profile, shocks are more likely to happen, which push
  more mass outwards and lead to more effective envelope expansions.
The broad existence of red profiles just represents small infall
  radius at the epoch and highly efficient envelope expansion.
Therefore by referring to underlying dynamic models, the presence
  of red profiles for nearly all impact parameter $b$ indicates
  that the central protostellar mass is very small (i.e.
  $\sim 0.106~M_\odot$ in Model IV and $\sim 0.012~M_\odot$
  in Model V) and the molecular cloud may form a brown dwarf
  at the centre.


\subsection{Effects of Optical Depth and Turbulence}

\begin{figure*}
\begin{center}
\epsfig{figure=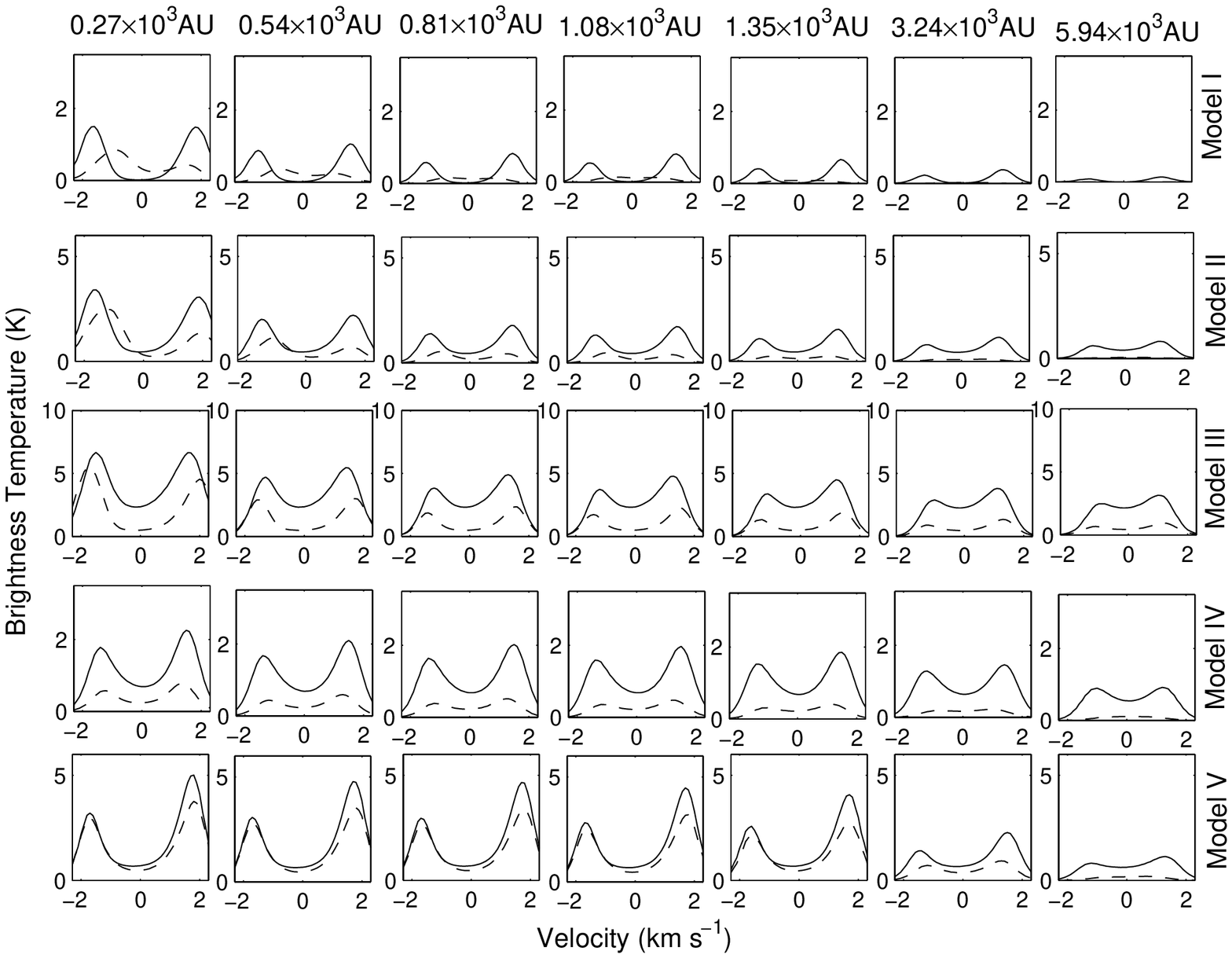,width=18.0cm,clip=}
\end{center}
\caption{Similar RATRAN computations for molecular spectral
  line profiles for five EECC dynamic Models I$-$V as in Fig.
  \ref{fig:spectra1}, but with intrinsic line broadening
  $\Delta u=1.5~{\rm km}~{\rm s}^{-1}$. This intrinsic line
  broadening contains both turbulence and thermal effects.
 \label{fig:spectra3}}
\end{figure*}

\begin{figure*}
\begin{center}
\epsfig{figure=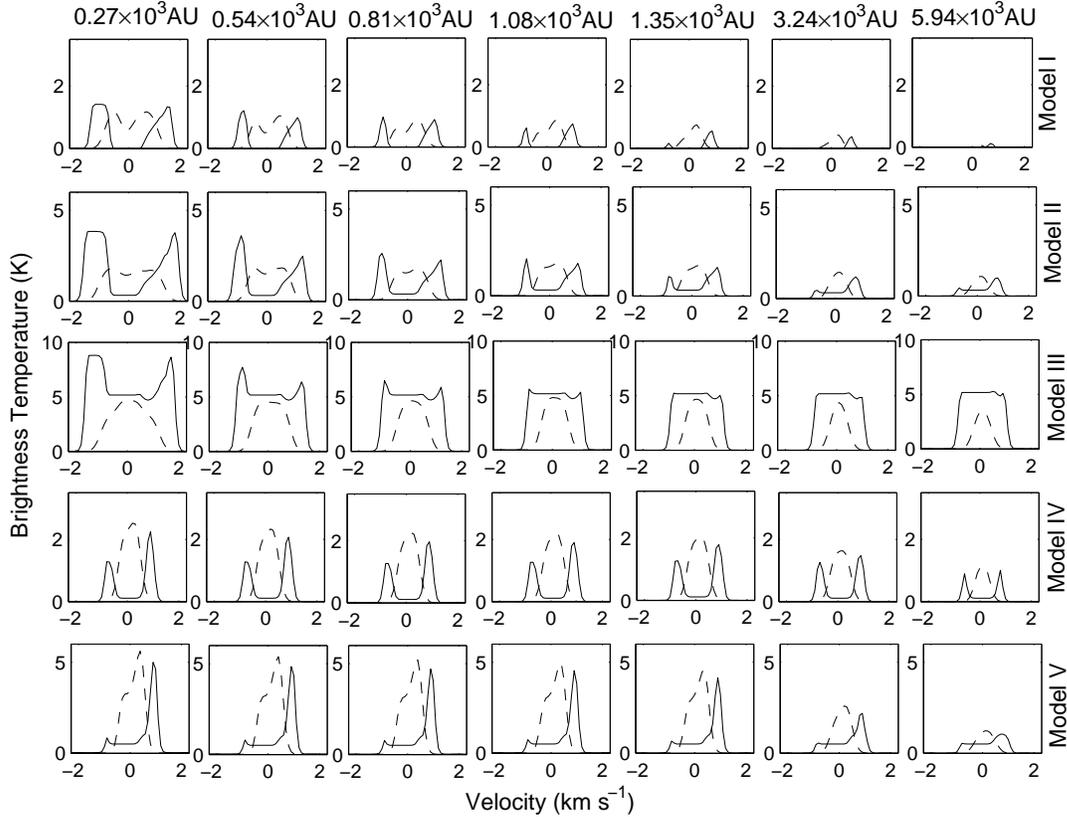,width=18.0cm,clip=}
\end{center}
\caption{Model calculations for spectral line profiles of
CO($2-1$) at 230.54~GHz (solid curves) and C$^{18}$O($1-0$) at
109.78~GHz (dashed curves) from individual pixels of spatially
resolved emission lines for five EECC dynamic Models I$-$V. The
ordinate is the brightness temperature in Kelvin and the abscissa
is the projected velocity component in the local standard of rest
(LSR) in unit of ${\rm km~s^{-1}}$. Rows from top to bottom show
molecular line profiles for different dynamic Models I$-$V,
respectively. Panels from left to right present LOS with different
impact parameter $b$, which are not uniformly chosen such that
spatial transitions between blue profiles and red profiles are
clearly identified for Models I, II and III. The intrinsic line
broadening is $\Delta u=0.5~{\rm km}~{\rm s}^{-1}$.
\label{fig:spectra4}}
\end{figure*}

\begin{figure*}
\begin{center}
\epsfig{figure=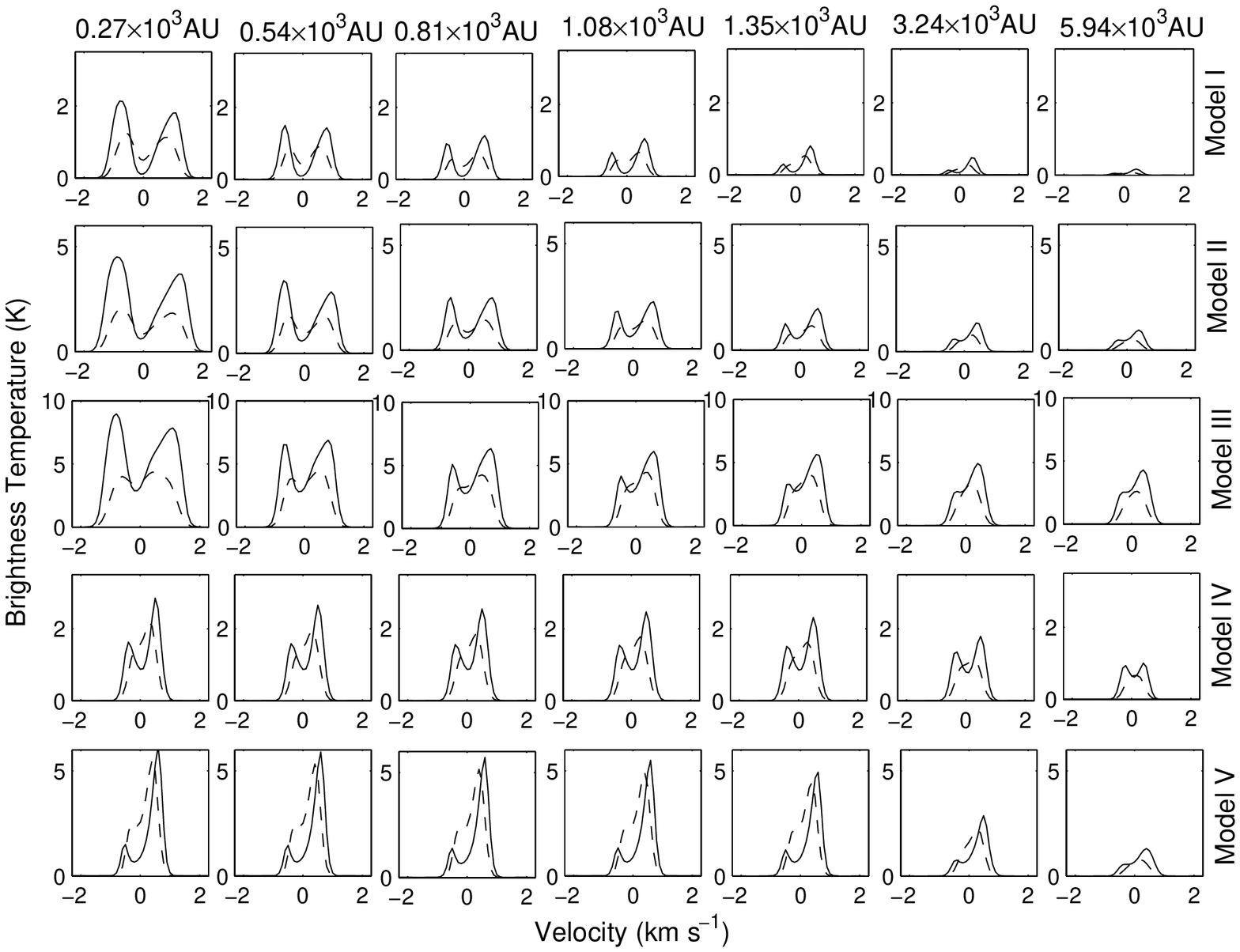,width=18.0cm,clip=}
\end{center}
\caption{Model calculations of spectral molecular line profiles of
CS($2-1$) at 97.98~GHz (solid curves) and N$_2$H$^+$($1-0$) at
93.13~GHz (dashed curves) from individual pixels of spatially
resolved emission lines for five EECC dynamic Models I$-$V. The
ordinate is the brightness temperature in Kelvin and the abscissa
is the projected velocity component in the local standard of rest
(LSR) in unit of ${\rm km~s^{-1}}$. Rows from top to bottom show
molecular line profiles for different dynamic Models I$-$V,
respectively. Panels from left to right present LOS  with
different impact parameter $b$, which are not uniformly chosen
such that transitions between blue profiles and red profiles are
clearly identified for Models I, II and III. The intrinsic line
broadening is $\Delta u=0.5~{\rm km}~{\rm s}^{-1}$.
\label{fig:spectra5}}
\end{figure*}

Proper optically thick conditions (i.e. absorption and scattering)
  in star-forming molecular clouds are responsible for such
  asymmetric spectroscopic signatures as in Figs.
  \ref{fig:spectra1}$-$\ref{fig:spectra3}. Both HCO$^+$ J$=1-0$ and
  J$=3-2$ transitions are deeply self-absorbed
  \citep[e.g.][]{tafalla2006}, while the J$=3-2$ line transition has
  a lower optical depth and source function because of lower level
  populations on J=3 and J=2 at such a cold environment with
  $T\sim 10~{\rm K}$.
This contrast causes two effects on molecular line profiles:
  first, the intensities of J$=3-2$ lines are weaker;
  and secondly, asymmetries of molecular line profiles are
  less apparent for J$=3-2$ transitions.
This comes to a widely tested conclusion that
  optically thick transition lines offer diagnosis for the
  large-scale thermal and dynamic structures of molecular clouds.
CO and CS transitions in Figs. \ref{fig:spectra4} and
  \ref{fig:spectra5} are also examples of optically thick
  transitions, which show distinct red profiles. In contrast,
  C$^{18}$O and N$_2$H$^+$ transitions are examples of optically
  thin transitions, which simply show single peak line profiles.

Turbulence in molecular clouds is another important aspect
   that affects molecular line profiles in a significant manner
   \citep[e.g.][]{arons1975,larson1981,LouRosner1986,zweibel1995,maclow1999}.
In this paper, the effect of turbulence is subsumed in the form
   of the intrinsic line broadening $\Delta u$, where another
   contribution to the broadening is thermal broadening of
   $\Delta u_{\rm the}\sim 0.1~{\rm km s^{-1}}$ for clouds
   with a temperature of $\sim 10~{\rm K}$.
We show radiative transfer results for a smaller or comparable
  turbulent broadening in Figs. \ref{fig:spectra1} and
  \ref{fig:spectra2}, and molecular line profiles for clouds under
  stronger turbulence are shown in Fig. \ref{fig:spectra3}.
Besides the increase of line width, line profile asymmetry
  decreases as a result of the enhanced turbulence.

\subsection{Star-Forming Molecular Clouds}

By fitting spatially resolved spectral line profiles to real
  molecular clouds, we may infer their underlying dynamic
  structures, and further estimate physical parameters (e.g.
  protostellar mass, central mass accretion rate,
  dynamic age etc.) of star-forming molecular cloud cores.
There are several aspects for the data fitting: (1) selection of
  self-similar EECC dynamic solutions, (2) adoption of proper physical
  scalings derived from empirical information and (3) choice of
  suitable turbulent broadening.
Additionally, sub-millimeter continuum observations can serve as a
  constraint on the radial profiles of density and temperature
  \citep[e.g.][]{adams1991,shirley2000,harvey2003} before fitting to molecular spectral
  emission line profiles.
For describing more realistic cloud situations, variations of
  molecular abundance ratio should be taken into account.
Star-forming molecular clouds L1517B \citep[e.g.][]{tafalla2006},
  L1544 \citep[e.g.][]{vandertak2005}, L1551NE \citep[e.g.][]{moriarty1995},
  L483 \citep[e.g.][Carolan et al. 2008]{park2000,tafalla2000}
  are among the good candidates of molecular clouds likely
  involving EECC dynamic motions.

In addition to red profiles for molecular lines, there are several
  other clues when applying EECC dynamic models to molecular clouds.
Global cloud systems have an estimated speed range of
  $\sim 0.1-1~{\rm km}~{\rm s}^{-1}$ for flows, which is smaller
  than bipolar outflows of typical speed $\gsim 10\ {\rm km}~{\rm s}^{-1}$.
Observed with sufficient spatial resolutions, molecular line
  profiles from clouds with EECC dynamics will maifest a circular symmetry;
  while bipolar outflows have line emissions with a bipolar asymmetry
  in spatial distributions, as we mentioned in Section 1.
Frequency resolution of about $0.1~{\rm km~s^{-1}}$ is needed to
  resolve spectroscopic signatures as shown in Figs.
  \ref{fig:spectra1} to \ref{fig:spectra3}.
High spatial resolution (e.g. 2'' or smaller for molecular clouds
  at $\sim 200$ pc) is also very important for detecting variations of
  molecular line profiles as a function of radius in cloud core,
  which helps to distinguish different underlying dynamic structures.

\section{Summary and Conclusions}

We invoke self-similar general polytropic EECC dynamic solutions
  without or with shocks to model the global evolution of a certain
  class of molecular clouds. By specifying relevant parameters
  plausibly estimated for molecular clouds, we illustrate several
  examples of general polytropic EECC cloud solutions. On the basis
  of these cloud solutions, we perform radiative transfer calculations
  to produce molecular line profiles to confirm the viability of EECC
  model framework.

Through extensive numerical explorations, we demonstrate that the
  widely observed `red profiles' in molecular emission spectral lines
  from star-forming clouds may well serve as important diagnostics for
  revealing the underlying EECC self-similar hydrodynamics in molecular
  clouds (Lou \& Shen 2004; Shen \& Lou 2004; Lou \& Gao 2006;
  Wang \& Lou 2008).
From the point of view of general polytropic hydrodynamics and
  radiative transfer, our explanation for the mystery of `red profiles'
  in emission spectral lines appears natural and physically sensible.
In particular, a molecular cloud characterized by an envelope
  expansion with a simultaneous central core collapse represents novel
  scenario.
Based on EECC solutions with or without shocks, optically thick
  molecular emission lines from gas clouds can show red profiles for
  all impact parameter $b$ of LOS from the cloud centres, or just
  outside a certain $b$ value (with the inner region showing blue
  profiles).
The optical depth of transition lines and turbulent broadening
  caused by micro gas motions will affect the appearance of red profiles.
Different from those of bipolar outflows, emission lines from clouds
  under EECC dynamics will grossly show circular spatial symmetry,
  and the flow speed is typically smaller (e.g.
  $\sim 0.1-1~{\rm km~s^{-1}}$).
By fitting spectral emission line profiles of certain dynamic
  models with observed emission lines from star-forming clouds, it
  is possible to resolve the dynamic structures of these molecular
  clouds. These processes will also give rise to more physical
  parameters of forming protostars and its cloud environment.


\section*{Acknowledgments}

This research was supported in part by Tsinghua
  Centre for Astrophysics (THCA), by the National Natural Science
  Foundation of China (NSFC) grants 10373009 and 10533020 at
  Tsinghua University, and by the Yangtze Endowment and the SRFDP
  20050003088 and 200800030071 at Tsinghua University.
The hospitality of Institut f\"ur Theoretische Physik und
  Astrophysik der Christian-Albrechts-Universit\"at Kiel Germany
  and of International Center for Relativistic Astrophysics
  Network (ICRANet) Pescara, Italy is gratefully acknowledged.
  YG thanks Y. Wu for assistance in running the RATRAN code.

\end{document}